\documentclass{amsart}
\usepackage{graphicx}
\usepackage{amssymb}
\usepackage{latexsym}
\usepackage{amsmath}
\usepackage{dsfont}
\usepackage{bbm}
\usepackage{scalerel,stackengine}
\stackMath
\newcommand\reallywidecheck[1]{%
\savestack{\tmpbox}{\stretchto{%
  \scaleto{%
    \scalerel*[\widthof{\ensuremath{#1}}]{\kern-.6pt\bigwedge\kern-.6pt}%
    {\rule[-\textheight/2]{1ex}{\textheight}}
  }{\textheight}%
}{0.5ex}}%
\stackon[1pt]{#1}{\scalebox{-1}{\tmpbox}}%
}
\usepackage{color}
\usepackage{xcolor}
\usepackage[breaklinks,colorlinks,backref]{hyperref}
\hypersetup{
    colorlinks, 
    linktoc=all, 
    linkcolor=red,  
  citecolor=hyptxt,
  urlcolor=blue}
  \hypersetup{
  citebordercolor=,
  filebordercolor=green,
  linkbordercolor=blue
}
\definecolor{hyptxt}{rgb}{0.7, 0.4, 0.9}
\usepackage{bibtopic}

\newcommand{\bra}{\begin{array}}
\newcommand{\era}{\end{array}}
\newcommand{\bqn}{\begin{eqnarray}}
\newcommand{\eqn}{\end{eqnarray}}
\newcommand\ben{\begin{enumerate}}
\newcommand\een{\end{enumerate}}
\newcommand\bei{\begin{itemize}}
\newcommand\ei{\end{itemize}}

\def\vk{\varkappa}
\def\lg{\langle}
\def\rg{\rangle}
\def\ud{\mathrm{d}}

\def\R{\mathbb{R}}

\def\C{\mathbb{C}}

\def\vr{\pmb{r}}
\def\vk{\pmb{k}}

\def\ii{\mathrm{i}}
\def\bu{\mathbbm{1}}
\numberwithin{equation}{section}

\date{}
\begin{document}

\markboth{Cohen-Tannoudji  \& Gazeau \&  Habonimana \& Shabani}
{Quantum fields with no Planck constant}

\title{Quantum models \`a la Gabor for space-time metric}

\author[Cohen-Tannoudji  \& Gazeau \&  Habonimana \& Shabani]{Gilles Cohen-Tannoudji$^a$, Jean-Pierre GAZEAU$^b$, \\ C\'elestin Habonimana$^c$, and Juma Shabani$^d$}

\address{$^a$ Laboratoire de recherche sur les sciences de la mati\`ere, LARSIM CEA, Universit\'e Paris-Saclay, F-91190 Saint-Aubin, France\\                                       
Gilles.Cohen-Tannoudji@cea.fr }

\address{$^b$ Universit\'e Paris Cit\'e, CNRS, Astroparticule et Cosmologie, F-75013 Paris, France  
\\ gazeau@apc.in2p3.fr}

\address{$^c$ Ecole Normale Sup\'erieure, Universit\'e du Burundi, Bujumbura, Burundi   
\\ habcelestin@yahoo.fr}

\address{$^d$  Universit\'e du Burundi, Bujumbura, Burundi   
\\ jushabani@yahoo.fr}

\maketitle

\begin{abstract}
As an extension of Gabor signal processing, the covariant Weyl-Heisenberg integral quantization is implemented to transform  functions on the eight-dimensional  phase space $\left(x,k\right)$ into Hilbertian operators. The $x=\left(x^{\mu}\right)$'s are space-time variables and the $k=\left(k^{\mu}\right)$'s are their conjugate wave vector-frequency variables. The procedure is first applied to the variables $\left(x,k\right)$ and produces canonically conjugate essentially  self-adjoint operators. It is next applied to the metric field $g_{\mu\nu}(x)$ of general relativity and yields regularised  semi-classical phase space portraits $\check{g}_{\mu\nu}(x)$. The latter give rise to modified tensor energy density.  Examples are given with the uniformly accelerated reference system  and  the Schwarzschild metric. Interesting probabilistic aspects 
 are discussed. 
\end{abstract}



\tableofcontents
\section{Introduction}
One of the difficulties to accept arbitrary change of space-time coordinates in general relativity lies in the primordial learning 
of our environment. For instance, in the case of the Schwarzschild metric, $\ud s^2 = (1-2m/r)\ud t^2  -(1-2m/r)^{-1}\ud r^2 -r^2\ud \theta^2 -r^2\sin^2\theta \ud\phi^2$, the radial coordinate  appears as  the most natural and easily understandable in terms of basic physics, namely a mass and a radius, while  numerous   coordinates (Gullstrand–Painlevé, Kruskal-Szekeres, Lemaître, Eddington–Finkelstein ...) used to examine the continuation of the Schwarzschild solution, are  not so immediate.  

From the early stages of  development  our senses undergo temporal and spatial dimensions of this environment. This training can be understood through the most basic physical laws like conservation of energy, momentum, angular momentum, through 
myriads of tiny collisions at all scales between any living matter and the rest of the world. In particular, the essentially Galilean perception of our tempo-spatial environment  might be  in mathematical terms compared with  Fourier reconstruction and analysis of tempo-spatial signals $s(x)\, , x:=(t,\vr)$ in the Hilbert space $L^2(\R^4, \ud^4x=\ud t\,\ud\vr)\equiv\mathcal{H}$.   
\begin{equation}
\label{fourspt}
\begin{split}
s(x)&= \frac{1}{4\pi^2}\int_{\R^4}\ud^4 k\, \hat{s}(k)\,e^{\ii k\cdot x}\, ,\quad k:=(\omega,\vk)\,,\quad k\cdot x:=\omega t-\vk\cdot\vr\,,\\  \hat{s}(k)&=  \frac{1}{4\pi^2}\int_{\R^4}\ud^4 x\, s(x)\,e^{-\ii x\cdot k}\, . 
\end{split}
\end{equation}
The minkowskian inner product $k\cdot x$ has been chosen for convenience, but,  in the present context,  the Fourier exponentials in the integrals \eqref{fourspt}  are not to be interpreted  as a travelling wave. They are just the basic ingredients of the Fourier transform for determining the frequency   $\omega$ and wave vector $\vk$ content of the signal $s(x)$.  The latter have inverse dimension of time and space vector  respectively, so no additional constants appear in the Fourier transform. Of course, they can appear in the expressions of the signals, and  they can  also be introduced into \eqref{fourspt} to give more physical flavor to the above formula in terms of energy and momentum. 

As is now well established in signal or image processing, Fourier techniques are advantageously replaced by Gabor reconstruction and analysis. One chooses a probe signal, or  \textit{window}, $\psi$, which is well localized in time-space   and frequency-wave-vector at once, and which is normalized in $\mathcal{H}$, $\Vert \psi \Vert^2 = \int_{\R^4} \ud^4 x\, \vert \psi(x) \vert^2$=1. This probe is then translated and modulated as
\begin{equation}
\label{JPGGabFam}
\psi(x) \rightarrow  \psi_{y, l}(x) = 
 e^{\mathsf{i}l\cdot x}\,\psi (x-y)\,.  
\end{equation}
We then get   the reconstruction of the signal $s(x)$ in $\mathcal{H}$ in  terms of its Gabor transform $S(x,k)$:
 \begin{equation}
\label{gabspt}
\begin{split}
s(x)&= \frac{1}{(2\pi)^4}\int_{\R^8}\ud^4 y\,\ud^4 l \, S(y,l)\, \psi_{y, l}(x),\\  S(x,k)&=  \langle \psi_{x,k} | s \rangle= \int_{\R^4}\ud^4 y\,\overline{\psi_{x, k}(y)}\,s(y) \, . 
\end{split}
\end{equation}
Like for Fourier conservation of the energy (in the signal sense) holds:
\begin{equation}
\int_{\R^4} \ud^4 x\, \vert s(x) \vert^2= \Vert s \Vert^2  = \frac{1}{(2\pi)^4}
\int_{\R^8} \ud^4 x\,\ud^4 k \, \vert S(x,k) \vert^2 := \Vert S \Vert^2\,.
\end{equation}
This \emph{Plancherel} formula results from the resolution of the identity operator $\mathbbm{1}$ in $\mathcal{H}$ provided by the continuous non-orthogonal family of \textit{coherent states} $\{\psi_{x,k}\,,\, (x,k)\in \mathbb{R}^8\}$ (overcompleteness):
\begin{equation}
\label{JPGgabres1}
\mathbbm{1} = \frac{1}{(2\pi)^4}
\int_{\R^8} \ud^4 x\,\ud^4 k \, \left|
\psi_{x,k}\right\rangle \left\langle \psi_{x,k} \right| \,.
\end{equation}
Like the Fourier transform is based on the one-dimensional unitary irreducible representations (UIR) of the abelian translation group in $\R^4$, the above Gabor transform is based on the infinite-dimensional von Neumann UIR of the Weyl-Heisenberg group $\mathrm{H}_{4}$
\cite{howe80}. This UIR is nothing else than the non-commutative extension of the translation group in $\R^8$ when the latter is combined with a one-parameter cocycle. Quantum mechanics for one Galilean elementary system  is itself based on the UIR of $\mathrm{H}_{3}$. 

In this article, we take advantage of the resolution of the identity   in the Hilbert space  of signals $\mathcal{H}$ to linearly map functions $f(x,k)$ on the phase space $\{(x,k)\}\sim \R^8$, and, more generally, distributions \cite{JPGber14},  to operators $A_f$ in $\mathcal{H}$. The procedure is called Weyl-Heisenberg covariant integral quantization, or more simply, Gabor quantization. According to the choice of the probe $\psi$ it can yield regularisations of the original 
$f$ through spectral properties of the operator $A_f$ and expected values  or \textit{lower symbols} or \textit{semi-classical phase space portraits} $\check{f}(x,k):= \left\langle \psi_{x,k} \right| A_f\left|
\psi_{x,k}\right\rangle$ of  $A_f$. The mathematical content of the present work is, of course, not original. It has been comprehensively developed over the past decades, with the noticeable contributions by Werner \cite{werner84}, Feichtinger and Kozek \cite{feichwer98},  Luef and Skrettingland \cite{luefskrett18}. 
In this context, we should also remind the pioneer works by Klauder \cite{klauder63,klauder64}, Berezin \cite{berezin75}, Daubechies and Grossmann \cite{daubgross80,daubgrossreig83} on the integral quantization based on standard coherent state, as  particular cases  of Gabor quantization with a quantum mechanics flavour. Concerning the genesis of specific aspects of our approach,  see for instance the recent \cite{JPGgaz18,JPGkono19,beczugamal20} \cite{gahumoze20,gazhabo20,gahumoze22} for more explanations on the motivations and illustrations.  The maps $f \mapsto A_f\mapsto \check{f}$ have  probabilistic features which have to be exploited within some informational interpretation and offer new perspectives, particularly in cosmology. Moreover, restricting the quantization to scalar fields $g(x)$, like metric fields of general relativity, leads to interesting outcomes. Of course the extension to tensorial fields is straightforward notwithstanding the increase of technicalities. 

The two following sections are heavily inspired from the recent \cite{gazhabo20}, and should be considered as a necessary reminder in view of extending our time-frequency ($\R^2$) approach to the time-space-frequency-wavevector $\R^8$.  
In Section \ref{JPGsigtoqu} is presented the integral quantization under its  general form.  
In Sections  \ref{JPGgabquant} and \ref{JPGwhiqg} we give an  outline  of the Gabor quantization and its generalisations for the simplest case of the time-frequency phase space $\{(t,\omega)\}\sim \R^2$. 
In Section \ref{infoentrop} we proceed with the analysis of the procedure in terms of observational and probabilistic aspects. 
In Section \ref{extentGR8} we extend this material to the eight-dimensional case for which the symmetry group is $\mathrm{H}_4$.  
However, for a sake of simplicity, we restrict our procedure  to the Gabor quantization based on the choice of probe functions $\psi$. In Section \ref{unac}, motivated by the 1935 Einstein-Rosen paper \cite{einstein-rosen35}  the Gabor quantization is applied to the  most elementary (and singular!) metric field of general relativity, namely the uniformly accelerated reference system. We show how the  regularising constant (introduced by hand by the authors) naturally emerges from our procedure. In Section \ref{szchw} we also consider another historical metric, the Schwarzschild solution, and show how the Gabor quantization and its phase space portraits lead to appealing regularisations, at the price of breaking the rotational symmetry. 
With  Section \ref{findis} we end this article by sketching some interesting perspectives.

\section{From signal analysis and reconstruction to integral quantization}
\label{JPGsigtoqu}

 Let $(X,\mu)$ be a measure space and $\mathcal{H}$ be a (separable) Hilbert space.   An operator-valued function
\begin{equation}
\label{}
X\ni x \mapsto \mathsf{M}(x)\ \mbox{acting in} \ \mathcal{H}\, , 
\end{equation}
resolves the identity operator $\mathbbm{1}$ in  $\mathcal{H}$ with respect the measure $\mu$ if 
\begin{equation}
\label{resUn}
\int_X \mathsf{M}(x)\, \mathrm{d} \mu(x)= \mathbbm{1} 
\end{equation}
holds in a weak sense, i.e., $\int_X \lg\psi|\mathsf{M}(x)|\psi^{\prime}\rg\, \mathrm{d} \mu(x)= \lg\psi|\psi^{\prime}\rg$ for any $\psi,\psi^{\prime}$ in the common domain of  the $\mathsf{M}(x)$'s, a.e. $x$. 

As we have in \eqref{gabspt},   \textit{analysis} and \textit{reconstruction} of a signal are described by the action of  \eqref{resUn} on a vector in $\mathcal{H}$
\begin{equation}
\label{ }
\mathcal{H}\ni | s\rangle \overset{\mbox{{\scriptsize reconstruction}}}{=} \int_X \overset{\mbox{{\scriptsize analysis}}}{\overbrace{\mathsf{M}(x)|s\rangle}}\, \mathrm{d} \mu(x)\,.
\end{equation}
 On the other hand,  \textit{integral quantization} is   the linear map of a function on $X$ to an operator in $\mathcal{H}$ which is defined by
\begin{equation}
\label{intqgen}
f(x) \mapsto \int_X f(x) \mathsf{M}(x)\, \mathrm{d} \mu(x)= A_f\, , \quad 1 \mapsto \mathbbm{1}\,.
\end{equation}
If the operators $\mathsf{M}(x)$ in \eqref{resUn}
are nonnegative and bounded, i.e., $\langle \phi | \mathsf{M}(x)|\phi\rangle  \geq 0$ for (almost) all $x\in X$ and all $\phi\in \mathcal{H}$, one says that  they form a (normalised) positive operator-valued measure (POVM) on $X$.  If they are further
 unit trace-class, i.e. $\mathrm{tr}(\mathsf{M}(x)) = 1$ for all $x\in X$, i.e., if the $\mathsf{M}(x)$'s are density operators, then the map 
\begin{equation}
\label{sempor}
f(x) \mapsto  \check{f}(x):= \mathrm{tr}(\mathsf{M}(x)A_f) = \int_X f(x^{\prime})\,\mathrm{tr}(\mathsf{M}(x)\mathsf{M}(x^{\prime}))\, \mathrm{d} \mu(x^{\prime})
\end{equation}
is a local averaging of the original $f(x)$ (which can very singular, like a Dirac !) with respect to the probability distribution on $X$,
\begin{equation}
x^{\prime} \mapsto \mathrm{tr}(\mathsf{M}(x)\mathsf{M}(x^{\prime}))\,. 
\end{equation}
This averaging, or semi-classical portrait of the operator $A_f$,  is in general a regularisation, depending of course on the topological nature of the measure space $(X,\mu)$ and the functional properties of the $\mathsf{M}(x)$'s.

Now,  consider a set of parameters $\boldsymbol{\kappa}$   and corresponding families of  POVM  $ \mathsf{M}_{\boldsymbol{\kappa}}(x)$ solving the identity
\begin{equation}
\label{resunH}
\int_X \mathsf{M}_{\boldsymbol{\kappa}}(x)\, \mathrm{d} \mu(x)= \mathbbm{1}\, ,  
\end{equation}
 One says that the \textit{classical limit} $f(x)$ holds at $\boldsymbol{\kappa}_0$ if 
\begin{equation}
\check{f}_{{\boldsymbol{\kappa}}}(x):= \int_X f(x^{\prime})\,\mathrm{tr}(\mathsf{M}_{\boldsymbol{\kappa}}(x)\mathsf{M}_{\boldsymbol{\kappa}}(x^{\prime}))\, \mathrm{d} \mu(x^{\prime}) \to f(x) \quad \mbox{as} \quad \boldsymbol{\kappa} \to \boldsymbol{\kappa}_0\, , 
\end{equation}
where the convergence $\check{f}\to f$ is defined in the sense of a certain topology. 

 Otherwise said, $\mathrm{tr}(\mathsf{M}_{\boldsymbol{\kappa}}(x)\mathsf{M}_{\boldsymbol{\kappa}}(x^{\prime}))$ tends to  
\begin{equation}
\mathrm{tr}(\mathsf{M}_{\boldsymbol{\kappa}}(x)\mathsf{M}_{\boldsymbol{\kappa}}(x^{\prime})) \to \delta_x(x^{\prime})
\end{equation}
where $\delta_x$ is a Dirac measure with respect to $\mu$,
\begin{equation}
\int_X f(x^{\prime}) \, \delta_x(x^{\prime})\, \mathrm{d} \mu(x^{\prime}) = f(x)\,.
\end{equation}
Actually,  nothing guarantees the existence of such  a limit on a general level. Nevertheless if the semi-classical $\check{f}_{{\boldsymbol{\kappa}}}$ might appear as  more realistic and easily handled than the original $f$, the question is to evaluate  the range of acceptability of the parameters $\boldsymbol{\kappa}$.

\section{Gabor quantization of functions of time-frequency variables}
\label{JPGgabquant}
In this section we present the material in the basic two-dimensional time-frequency case in order to progressively make the reader familiar to our quantization method. 

\subsection{Gabor quantization}

In order to manage time-frequency functions $f(b,\omega)$, we first start from the resolution of the identity provided by the Gabor POVM. The latter is built as follows.   
\begin{equation}
\label{ }
\mathbbm{1}= \int_{\mathbb{R}^2}\frac{\mathrm{d} b\,\mathrm{d} \omega}{2\pi} |\psi_{b \omega}\rangle \langle \psi_{b \omega}| \, ,\quad  \langle \psi_{b \omega}| \psi_{b^{\prime} \omega^{\prime}}\rangle\neq \delta(b-b^{\prime})\, \delta(\omega-\omega^{\prime})
\end{equation}
 where the functions 
 \begin{equation}
\label{gabom}
\langle\delta_t| \psi_{b \omega}\rangle := e^{\mathsf{i} \omega t} \psi(t-b)
\end{equation}
 are the modulated-transported unit-norm probe-vectors in $L^2(\mathbb{R},\mathrm{d} t)$
and where $\mathbb{R} \supset\Delta \mapsto  \int_{\Delta}\frac{\mathrm{d} b\,\mathrm{d} \omega}{2\pi} |\psi_{b \omega}\rangle \langle \psi_{b \omega}|$ is the corresponding normalised POVM on the plane. From \eqref{intqgen} 
the quantization of $f(b,\omega)$ is given by:
\begin{equation}
\label{JPGgabq1}
f\mapsto A_f = \int_{\mathbb{R}^2}\frac{\mathrm{d} b\,\mathrm{d} \omega}{2\pi}\, f(b,\omega)\,  |\psi_{b \omega}\rangle \langle \psi_{b \omega}| \, .
\end{equation}
 The latter maps a function (or tempered distribution) $f(b,\omega)$ on the time-frequency plane to the integral operator $A_f$ acting  in the Hilbert space of finite-energy signals as
\begin{equation}
\label{gabAfdef}
(A_f s)(t) = \langle \delta_t  | A_f | s \rangle = \int_{-\infty}^{+\infty}\mathrm{d} t^{\prime}\, \mathcal{A}_f(t,t^{\prime})\, s(t^{\prime})\, , 
\end{equation}
with integral kernel given by
\begin{equation}
\label{gabqker}
 \mathcal{A}_f(t,t^{\prime})= \frac{1}{\sqrt{2\pi}}\int_{-\infty}^{+\infty}\mathrm{d} b\, \widehat{f}_{\omega}(b,t^{\prime}-t)\, \psi(t-b)\,\overline{\psi(t^{\prime}-b)}
\end{equation}
Here $\widehat{f}_{\omega}(b,y)$ is the partial Fourier transform with respect to the variable $\omega$:
\begin{equation}
\label{parFT}
\widehat{f}_{\omega}(b,y)= \frac{1}{\sqrt{2\pi}}\int_{-\infty}^{+\infty}\mathrm{d} \omega\, f(b,\omega) \, e^{-\mathsf{i} \omega y}\,.
\end{equation}
The outcome of the construction of the operator $A_f$ through \eqref{gabAfdef} should be analysed from the quantum measurement viewpoint.   
Within the framework of  quantum physics, a physically relevant operator of the above form $A_f$ is a self-adjoint operator whose expectation value is the ``unsharp'' representation \cite{holevo11,gazeau-heller15,busch16} 
\begin{align}
	\label{measexpect}
	\mathrm{tr}\left(\rho_m A_f\right) =  \int_{X}f(x)\,\mathrm{tr}(\rho_m\rho(x))\,\mathrm{d}\nu(x)\, , 
\end{align}
where $\rho_m$ is a density operator ($\sim$ mixed quantum state) describing the way the physical system under observation has been prepared. In \eqref{JPGgabq1} a real function $f(b,\omega)$ is mapped to a symmetric operator and if $f$ is semi-bounded  then  $A_f$ is self-adjoint through the Friedrichs extension \cite{akhglaz81} of its associated semi-bounded quadratic form. For unbounded real $f$, self-adjointness is not guarantee and should be individually investigated.  

  Let us go through some specific situations. The Gabor quantization of separable functions $f(b,\omega)= u(b)v(\omega)$ yields the integral kernel
\begin{equation}
 \mathcal{A}_{uv}(t,t^{\prime})= \frac{\hat v(t^{\prime}-t)}{\sqrt{2\pi}}\int_{-\infty}^{+\infty}\mathrm{d} b\,u(b)\, \psi(t-b)\,\overline{\psi(t^{\prime}-b)}\,,
\end{equation}
where $\hat v$ is the Fourier transform of $v$. 
Therefore the action on a signal $s(t)$ reads as the combination of convolution and multiplication
\begin{equation}
\label{gabquv}
(A_{uv} s)(t) =\frac{1}{\sqrt{2\pi}} \int_{-\infty}^{+\infty}\mathrm{d} b \, \psi(b) \,u(t-b) \, \left( \overline{\tilde\psi}_b \hat{\tilde v}\ast s\right)(t)\, , \quad  \psi_b(t):=\psi(t-b)\,,
\end{equation}
where $\tilde{\psi}(t):= \psi(-t)$.
In the monovariable case $f(b,\omega)= u(b)$ the integral kernel \eqref{gabqker} reads as 
\begin{equation}
\label{gabquk}
 \mathcal{A}_{u}(t,t^{\prime})= \delta(t^{\prime}-t)\int_{-\infty}^{+\infty}\mathrm{d} b\,u(b)\, \vert\psi(t-b)\vert^2= \delta(t^{\prime}-t)\, \left(\vert \psi\vert^2\ast u\right)(t)\,,
 \end{equation}
 and one gets the multiplication operator
 \begin{equation}
\label{gabqu} (A_{u(b)} s)(t)= \left(u\ast\vert \psi\vert^2\right)(t)\,s(t)\,.
\end{equation}
For  the other monovariable case $f(b,\omega)= v(\omega)$ one gets  the integral kernel 
\begin{equation}
\label{JPGgabqv}
 \mathcal{A}_{v(\omega)}(t,t^{\prime})= \frac{\hat v(t^{\prime}-t)}{\sqrt{2\pi}}\int_{-\infty}^{+\infty}\mathrm{d} b\, \psi(b)\,\overline{\tilde\psi}(t-t^{\prime}-b)=  \frac{\hat v(t^{\prime}-t)}{\sqrt{2\pi}}\, R_{\psi\psi}(t-t^{\prime})
 \end{equation}
 where 
 \begin{equation}
\label{autocorrel}
R_{\psi\psi}(t):= \int_{-\infty}^{+\infty}\mathrm{d} t^{\prime}\,\psi(t^{\prime})\,\overline{\psi(t^{\prime}-t)}=\left(\psi \ast \overline{\tilde{\psi}}\right)(t)=\overline{\widetilde{R}}_{\psi\psi}(t)
\end{equation}
  is the autocorrelation of the probe,  i.e. the  correlation of the probe with a delayed copy of itself as a function of delay.
   Note that 
\begin{equation}
\frac{1}{\sqrt{2\pi}}R_{\psi\psi}(t)= \mathcal{F}^{-1}\left[\vert\hat \psi\vert^2\right](t)\,.
\end{equation}
We eventually get the  convolution operator  acting on the signal:
 \begin{equation}
\label{ }
 (A_{v(\omega)} s)(t)= \frac{1}{\sqrt{2\pi}} \left[\left(R_{\psi\psi}\hat{\tilde v}\right)\ast s\right](t)
\end{equation}

The application of these formulae to time and frequency variables yields the centered time and frequency operators:
\begin{align}
\label{JPGgabqt}
    A_b &= T  -\lg T\rg_{\psi} \,\mathbbm{1} \,, \quad (Ts)(t)= ts(t)\,, \\
 \label{JPGgabqf}     A_\omega &= \Omega  -\lg \Omega\rg_{\psi}\, \mathbbm{1} \,,\quad (\Omega s)(t)= -\mathsf{i} \,\partial_{t}s(t)\,,
\end{align}
where $\lg A\rg_{\psi}$ is the mean value of the operator $A$ in the state $\psi$, as it is usually denoted within the quantum framework. The choice of an even probe $\psi$ allows to get rid of these constants. It is the case with the standard choice of  the normalised centred  Gaussian with width $\sigma$ 
\begin{equation}
\label{gaussian}
\psi(t) \equiv G_{\sigma}(t)=\frac{1}{\pi^{1/4}\sqrt{\sigma}}e^{-\dfrac{t^2}{2\sigma^2}}\,.
\end{equation} 
Note that its autocorrelation is also a (not normalised) Gaussian: 
\begin{equation}
\label{JPGautocorG}
R_{G_{\sigma}G_{\sigma}}(t)= e^{-\frac{t^2}{4\sigma^2}}\,.
\end{equation}

The operators $T$ and $\Omega$, like  $A_b$ and $A_\omega$, are essentially self-adjoint and obey the ``canonical" commutation rule (no $\hbar$ here!):
\begin{equation}
T\,\Omega -\Omega\,T \equiv [T,\Omega] = \mathsf{i} \,\mathbbm{1}\,,
\end{equation}
with its immediate Fourier uncertainty consequence 
\begin{equation}
\label{ }
\Delta_s T\, \Delta_s \Omega\geq \frac{1}{2}\, , \quad \Delta_s A:= \sqrt{\langle s| A^2|s\rangle  - (\langle s|A|s\rangle)^2}
\end{equation}
  Now, one should keep in mind that the CCR  $[A,B] = \mathrm{i} \mathbbm{1}$  for a  self-adjoint ($A$, $B$) pair, with common domain, holds true only if both have continuous spectrum $(-\infty,+\infty)$. The expression of the CCR in terms of the respective unitary operators reads as
  \begin{equation}
\label{JPGWeylrel}
 e^{\mathrm{i} \sigma \Omega}\,e^{\mathrm{i} \tau T}= e^{\mathrm{i}\sigma \tau} e^{\mathrm{i}\tau T}\, e^{\mathrm{i} \sigma \Omega}\,,  \quad \mbox{(Weyl relations)}\,.
\end{equation}
 von Neumann proved (1931) 
  that up to multiplicity and unitary equivalence the Weyl relations have only one solution (see \cite{JPGreedsimon75} for the proof).
In the present formalism, time is elevated to the status of a quantum observable. This  is absolutely not the case for the time operator of quantum mechanics 
 (see for instance the review \cite{timeQM}). Indeed, in quantum mechanics the Hamiltonian  is bounded below and so  the quantum time, although  symmetric, is not self-adjoint and has no self-adjoint extension.

Concerning the square of time and frequency variables we obtain:
\begin{align}
\label{JPGgab2qt2}
    A_{b^2} &= (T  -\lg T\rg_{\psi})^2 +  \Delta^2_\psi T\,\mathbbm{1} \,, \\
 \label{JPGgabqf}     A_{\omega^2} &= (\Omega  -\lg \Omega\rg_{\psi})^2 +  \Delta^2_\psi \Omega\,\mathbbm{1}  \,.
\end{align}
With the choice of Gaussian probe \eqref{gaussian} these operators read: 
\begin{equation}
\label{gauss2qt2}
    A_{b^2} = T^2 + \frac{\sigma^2}{2}\, \mathbbm{1}\,, \quad
A_{\omega^2} = \Omega^2 + \frac{\sigma^2}{2}\, \mathbbm{1} \,.
\end{equation}

\subsection{Gabor semi-classical portrait }

The Gabor semi-classical  phase-space portrait of $A_f$ is given by
\begin{equation}
\label{JPGgabsc1} \check{f}(b,\omega)=  \int_{\mathbb{R}^2}\frac{\mathrm{d} b^{\prime}\,\mathrm{d} \omega^{\prime}}{2\pi}\, f(b^{\prime},\omega^{\prime})\,  \vert\langle \psi_{b \omega}|\psi_{b^{\prime} \omega^{\prime}}\rangle\vert^2\,.
\end{equation}
The overlap $\langle \psi_{b \omega}|\psi_{b^{\prime} \omega^{\prime}}\rangle$ is given by
\begin{equation}
\label{ovlap}
\begin{split}
\langle \psi_{b \omega}|\psi_{b^{\prime} \omega^{\prime}}\rangle&= \int_{\R}\ud t \, e^{-\ii (\omega^{\prime}-\omega)t}\,\overline{\psi(t-b^{\prime})}\,\psi(t-b)\\
&= 
e^{-\ii(\omega^{\prime}-\omega)b}\left(\mathcal{F}\left[\mathtt{t}_{b^{\prime}-b}\bar{\psi}\right]\ast \mathcal{F}\left[\psi\right]\right)(\omega^{\prime}-\omega)\,,
\end{split}
\end{equation}
where $\mathcal{F}[f]:=\hat{f}$ (Fourier transform), $(\mathtt{t}_bf)(t):= f(t-b)$,  and  
$$(b^{\prime}, \omega^{\prime}) \mapsto \vert\langle \psi_{b \omega}|\psi_{b^{\prime} \omega^{\prime}}\rangle\vert^2/2\pi$$
 is a probability distribution on the phase space equipped with the Lebesgue measure $\mathrm{d} b^{\prime}\,\mathrm{d} \omega^{\prime}$. Hence the regularity properties of $\check{f}$ depend on  those of the probe.  The function $\check{f}(b,\omega)$ is an elaborate combination a partial Fourier transform and multi-convolutions:
\begin{equation}
\label{lowsf}
\check{f}(b,\omega)= \frac{1}{\sqrt{2\pi}} \int_{\R}\ud \tau\, e^{\ii\omega\tau}\int_{\R}\ud b^{\prime}\, \widehat{f}_{\omega}(b-b^{\prime},\tau)\,\left(\left(\overline{\tilde{\psi}}\,\widetilde{\mathtt{t}_{b^{\prime}}\psi}\right)\ast\left(\overline{\mathtt{t}_{-b^{\prime}}\psi}\,\psi\right)\right)(\tau)\,. 
\end{equation}
In  the monovariable case $f(b,\omega)= u(b)$ this formula simplifies to
\begin{equation}
\label{lowu}
\check{u}(b,\omega)\equiv \check{u}(b)= \left(u\ast\left(\vert\psi\vert^2\ast\vert\widetilde{\psi}\vert^2\right)\right)(b)=  \left(u\ast R_{\vert\psi\vert^2\vert\psi\vert^2}\right)(b)\,. 
\end{equation}
where we note the appearance of the autocorrelation of the probability distribution $t\mapsto \vert\psi(t)\vert^2$ on the temporal axis. For the simplest cases time and time squared we get
\begin{equation}
\label{lowtt2}
\check{b}= b - \lg b\rg_{R_{\vert\psi\vert^2\vert\psi\vert^2}}\, , \quad \check{b^2}= \left(b- \lg b\rg_{R_{\vert\psi\vert^2\vert\psi\vert^2}}\right)^2  + \sigma^2_{R_{\vert\psi\vert^2\vert\psi\vert^2}}(b)\,, 
\end{equation}
where $\lg s\rg_{p}$ stands for the expected value of the signal $s$ with respect to the probability distribution $p$, and $\sigma^2_{p}(s(t))$ is its variance  with respect to  $p$.
An analogous formula holds (in the Fourier side) for  the monovariable case $f(b,\omega)= v(\omega)$:
\begin{equation}
\label{lowv}
\check{v}(b,\omega)\equiv \check{v}(\omega)= \overline{\mathcal{F}}\left[\hat{v}\left(\overline{\tilde{\psi}}\ast\psi\right)^2\right](\omega)\,,\quad  \overline{\mathcal{F}}:= \mathcal{F}^{-1}\,. 
\end{equation}

With the choice of Gaussian probe \eqref{gaussian} the overlap \eqref{ovlap} is given by
\begin{equation}
\label{gaussovl}
\lg G_{\sigma,b^{\prime},\omega^{\prime}}| G_{\sigma,b,\omega}\rg= e^{-\ii(\omega^{\prime}-\omega)\dfrac{b+b^{\prime}}{2}} \,e^{-\dfrac{(b-b^{\prime})^2}{4\sigma^2}}\,e^{-\dfrac{\sigma^2(\omega-\omega^{\prime})^2}{4}}\, , 
\end{equation}
and the semi-classical portrait of the operator $A_f$ is  the double Gaussian convolution: 
\begin{equation}
\label{gaussc}
\check{f} (b,\omega)= \int_{\mathbb{R}^2}\frac{\mathrm{d} b^{\prime}\,\mathrm{d} \omega^{\prime}}{2\pi}\, f(b^{\prime},\omega^{\prime})\, e^{-\dfrac{(b-b^{\prime})^2}{2\sigma^2}}\,e^{-\dfrac{\sigma^2(\omega-\omega^{\prime})^2}{2}}\,.
\end{equation}
As a consequence we observe that  no classical limit holds at $\sigma \to 0$ or $\sigma \to \infty$. This is just an illustration of the time-frequency uncertainty  principle.
Separable functions $f(b,\omega) = u(b)v(\omega)$ remain separable. From Eq.\,\eqref{lowu} and the convolution of two Gaussian, one finds: 
\begin{equation}
\label{lowsep}
\check{f} (b,\omega) = \left(u\ast G^2_{\sqrt{2}\sigma}\right)(b)\,\left(v\ast G^2_{\sqrt{2}/\sigma}\right)(\omega)\,,
\end{equation}
and in particular
\begin{equation}
\label{lowsepbom}
\check{u} (b) = \left(u\ast G^2_{\sqrt{2}\sigma}\right)(b)\,, \quad
\check{v} (\omega) = \left(v\ast G^2_{\sqrt{2}/\sigma}\right)(\omega)\,.
\end{equation}
Here the classical limit exists separately, and the convergence is simple at least for regular enough $u$ and $v$:
\begin{equation}
\label{sepclalim}
\check{u} (b)\underset{\sigma\to 0}{\to} u(b)\, , \quad \check{v} (\omega)\underset{\sigma\to \infty}{\to} v(\omega)\,. 
\end{equation}
For the simplest cases, 
\begin{align}
\label{chbb2}
   \check{b} &= b\, , \quad  \check{b^2}= b^2 + \sigma^2\,,  \\
  \label{chomom2}   \check{\omega} &= \omega\, , \quad  \check{\omega^2}= \omega^2 +   \frac{1}{\sigma^2}\,.
\end{align}

\section{Beyond Gabor: Weyl-Heisenberg covariant  integral  quantization of the time-frequency plane}
\label{JPGwhiqg}

Gabor signal analysis and quantization are the simplest ones among a vast amount of possibilities, all of them being based on the unitary dual of the Weyl-Heisenberg group. 
Let us remind the most important  features of this group  that we use in our approach to quantization. More details are given in the pedagogical 
\cite{JPGgaz18}. 

We recognize in the construction of the Gabor  family \eqref{JPGGabFam} the combined actions of the two unitary operators introduced in \eqref{JPGWeylrel}, with respective generators the self-adjoint time and frequency operators
\begin{equation}
\label{JPGGabWH1}
L^2(\mathbb{R}, \mathrm{d}t) \ni \psi(t) \mapsto  \psi_{b,\omega}(t)=  \left(e^{\mathrm{i} \omega T} e^{-\mathrm{i} b\Omega} \psi\right)(t)\, ,
\end{equation}
Two alternative forms of the action \eqref{JPGGabWH1} are provided by the Weyl formulae \eqref{JPGWeylrel} combined with the Baker-Campbell-Hausdorff formula:
\begin{equation}
\label{JPGGabWH2}
\psi_{b,\omega}(t) =e^{\mathrm{i} b\omega} \left( e^{-\mathrm{i} b\Omega} e^{\mathrm{i} \omega T}\psi\right)(t)= e^{\mathrm{i} \frac{b\omega}{2}}\left(e^{\mathrm{i} ( \omega T -b \Omega )}\psi\right)(t)
 \, .
\end{equation}
In the above appears  the Weyl 
or displacement operator up to a phase factor 
\begin{equation}
\label{JPGGabWH3}
e^{\mathrm{i} ( \omega T -b \Omega)}= \mathcal{D}(b,\omega)\, , \quad \psi_{b,\omega}(t) = e^{\mathrm{i} \frac{b\omega}{2}}  \left(\mathcal{D}(b,\omega)\psi\right)(t)
 \, .
\end{equation}
The appearance of this phase factor like that one appearing in the composition formula \eqref{JPGWeylrel} indicates that 
the map $(b,\omega) \mapsto \mathcal{D}(b,\omega)$ is a \textit{projective}
unitary representation of the time-frequency abelian plane. Dealing with a true representation necessitates the introduction of a third degree of freedom to account for this phase factor.  Hence we are led to work with the Weyl-Heisenberg group $\mathrm{G}_{\rm WH}$.   
\begin{equation}
\label{JPGWHgroupGab}
 G_{\mathrm{WH}}= \{ g = (\varsigma , b,\omega)\, , \, \varsigma \in {\mathbb R}\,,\,  (b,\omega)  \in  {\mathbb R}^{2}\} \, ,
\end{equation} 
with neutral element: $(0,0,0)$, and
\begin{equation}
\label{JPGWHGablaw}
g_{1}g_{2} = \left(\varsigma_{1} + \varsigma_{2} + \frac 1{2} (\omega_{1}b_{2} - \omega_2b_{1})\, ,   
        b_1 + b_2   ,\; \omega_{1}+\omega_{2}\right) \, , \quad g^{-1}= (-\varsigma,-b, -\omega)\,. 
\end{equation}
 The Weyl-Heisenberg group symmetry underlying the Gabor transform 
 is understood through its unitary irreducible representation (UIR). As a result of the von-Neumann uniqueness theorem, any infinite-dimensional UIR, $U$, of $\mathrm{G}_{\rm WH}$ is characterized by a real number $\lambda \neq 0$
(there is also the degenerate, one-dimensional, UIR corresponding to $\lambda = 0$).
 If the  Hilbert space carrying the UIR is the space of finite-energy signals $\mathcal{H} = L^{2}({\mathbb R},\mathrm{d}t)$, the representation operators  are defined by the action similar to \eqref{JPGGabWH1} (with the choice $\lambda = 1$) and completed with a phase factor:
\begin{equation}
\label{WHUIREP1}
U(\varsigma , b, \omega)= e^{\mathrm{i} \varsigma}
    e^{-\mathrm{i} \omega b/2} \, e^{\mathrm{i} \omega T}\, e^{-\mathrm{i} b \Omega} = e^{\mathrm{i} \varsigma}\, \mathcal{D}(b,\omega)\,. 
\end{equation}
 We now pick a bounded traceclass operator  $\mathfrak{Q}_0$ on $\mathcal{H}$ with
 \begin{equation}
\label{trQ01}
\mathrm{tr}\,\mathfrak{Q}_0= 1\, .
\end{equation}
 Its  unitary Weyl-Heisenberg transport yields the continuous family of 
bounded traceclass operators
\begin{equation}
\label{ }
\mathfrak{Q}(b,\omega)= U(\varsigma,b,\omega)\mathfrak{Q}_0 U(\varsigma,b,\omega)^{\dag}=  \mathcal{D}(b,\omega)\mathfrak{Q}_0  \mathcal{D}(b,\omega)^{\dag}\,.
\end{equation}
Applying the  Schur  Lemma  to the irreducible projective unitary representation $(b,\omega) \mapsto \mathcal{D}(b,\omega)$ allows to prove 
 the resolution of the identity obeyed by the operator-valued function  $\mathfrak{Q}(b,\omega)$ on the time-frequency plane 
\begin{equation}
\label{ }
\int_{\mathbb{R}^2}  \mathfrak{Q}(b,\omega)\, \frac{\mathrm{d} b \,\mathrm{d} \omega}{2\pi} = \mathbbm{1}\,. 
\end{equation}
It ensues the Weyl-Heisenberg integral quantization in its most general formulation:
\begin{equation}  
\label{JPGWHCIQ1}
f(b,\omega) \mapsto A_f= \int_{\mathbb{R}^2} f(b,\omega)\, \mathfrak{Q}(b,\omega)\, \frac{\mathrm{d} b\, \mathrm{d}\omega}{2\pi}\,.
\end{equation}
If $\mathfrak{Q}_0$ is symmetric,  a real function $f(b,\omega)$ is mapped to a symmetric operator $A_f$. Moreover, if   $\mathfrak{Q}_0$ is  non-negative, i.e., is a density operator, then  a real semi-bounded  function $f(b,\omega)$ is mapped to a self-adjoint operator $A_f$ as it was already stated about the Gabor quantization \eqref{gabAfdef}.

Due to the  homogeneity of the phase space, i.e. the choice of the origin is arbitrary, we have translational covariance in the sense that the quantization of the translated of $f$ is unitarily equivalent to the quantization of $f$: 
\begin{equation}
\label{covtrans1}\begin{split}
&U(\varsigma,b_0,\omega_0)\,A_f \,U(\varsigma,b_0,\omega_0)^{\dag}= A_{\mathcal{T}(b_0,\omega_0)f}\, , \\ &\mbox{where}\ \left(\mathcal{T}(b_0,\omega_0)f\right)(b,\omega):= f\left(b-b_0,\omega-\omega_0\right) \,. 
\end{split}
\end{equation} 

Note that the Gabor quantization \eqref{JPGgabq1} corresponds to the choice of the orthogonal projector $\mathfrak{Q}_0= |\psi\rangle\langle \psi|$. 
 
 There exists an  equivalent form to  the quantization \eqref{JPGWHCIQ1}. This form is expressed in terms of the displacement operator $\mathcal{D}$, the symplectic Fourier transform of $f(b,\omega)$,
\begin{equation}
\label{JPGSFT}
\mathfrak{f}_s[f](b,\omega):= \int_{\mathbb{R}^2}e^{-\mathrm{i} (b\omega^{\prime}-b^{\prime}\omega)}\, f(b^{\prime},\omega^{\prime})\,\frac{\mathrm{d} b^{\prime}\,\mathrm{b} \omega^{\prime} }{2\pi}\,, \quad  \mathfrak{f}_s^2= \mathbbm{1}\,,
\end{equation}
 and the Weyl transform of the operator $\mathfrak{Q}_0$ defined as the ``apodization'' function on the time-frequency plane
 \begin{equation}
\label{JPGWTQ}
\Pi(b,\omega) := \mathrm{tr}\left(\mathcal{D}(-b,-\omega)\mathfrak{Q}_0 \right)\, , \quad \Pi(0,0) = 1\,.
\end{equation}
The  inverse of the latter transform  is given by
\begin{equation}
\label{invweylQ_0}
\mathfrak{Q}_0= \int_{\mathbb{R}^2}  \mathcal{D}(b,\omega)\, \Pi(b,\omega) \,\frac{\mathrm{d} b\,\mathrm{d}\omega }{2\pi }\,,
\end{equation}
which means that $\mathfrak{Q}_0$ is the quantization of $f(b,\omega)=2\pi \delta(b,\omega)$ such that $\overline{\mathfrak{f}_s}[f](b,\omega)= 1$.

One obtains
 \begin{equation} 
A_f= \int_{\mathbb{R}^2}  \mathcal{D}(b,\omega)\,  \overline{\mathfrak{f}_s}[f](b,\omega)\, \Pi(b,\omega) \,\frac{\mathrm{d} b\,\mathrm{d}\omega }{2\pi } \,.
\end{equation}
From this form is obtained the action $A_f$ as the  integral operator  
\begin{equation}
\label{Afint}
\mathcal{L}^2(\R,\ud t) \ni s(t) \mapsto (A_f s)(t) = \int_{-\infty}^{+\infty}\mathrm{d} t^{\prime}\, \mathcal{A}_f(t,t^{\prime})\, s(t^{\prime})\, , 
\end{equation}
with  kernel given by
\begin{equation}
\label{AfintK}
 \mathcal{A}_f(t,t^{\prime})= \frac{1}{2\pi}\int_{-\infty}^{+\infty}\mathrm{d} b\, \widehat{f}_{\omega}(b,t^{\prime}-t)\, \widehat{\Pi}_{\omega}\left(t-t^{\prime},b- \frac{t+t^{\prime}}{2}\right)\,,
\end{equation}
where the partial Fourier transforms $\widehat{f}_{\omega}$ and $\widehat{\Pi}_{\omega}$ are defined in \eqref{parFT}.

The semi-classical portrait \eqref{sempor} of $A_f$ now reads:
\begin{equation}
\label{semporwh}
\check{f}(b,\omega) =\mathrm{tr}( \mathfrak{Q}(b,\omega)A_f) = \int_\mathrm{\R^2} f(b^{\prime},\omega^{\prime})\, \mathrm{tr}\left( \mathfrak{Q}(b,\omega)\mathfrak{Q}(b^{\prime},\omega^{\prime})\right)\, \frac{\mathrm{d} b^{\prime}\, \mathrm{d}\omega^{\prime}}{2\pi}\, .
\end{equation}
 Equivalently:
\begin{equation} 
\label{lowsymfsfs} 
\begin{split}
\check{f}(b,\omega)  &= \int_{\mathbb{R}^2}  \mathfrak{f}_s\left[\Pi\,\widetilde\Pi\right](b-b^{\prime}, \omega-\omega^{\prime})\, f(b^{\prime},\omega^{\prime}) \,\frac{\mathrm{d} b^{\prime}\,\mathrm{d} \omega^{\prime}}{2\pi}  \\
&=\int_{\mathbb{R}^2}  \mathfrak{f}_s\left[\Pi\right]\ast\mathfrak{f}_s\left[\widetilde\Pi\right](b- b^{\prime}, \omega-\omega^{\prime})\, f(b^{\prime},\omega^{\prime}) \,\frac{\mathrm{d} b^{\prime}\,\mathrm{d} \omega^{\prime}}{4\pi^2} \,,
\end{split}
\end{equation}
where $\widetilde\Pi(b,\omega)= \Pi(-b,-\omega)$.

\section{Probabilistic  aspects}
\label{infoentrop}
Let us  develop the probabilistic content of the quantization and smoothing procedure described in the previous section. 

With a true probabilistic content, i.e., with a  choice of operator  $\mathfrak{Q}_0$, or, equivalently, with a choice of  function $\Pi$ in \eqref{JPGWTQ},  such that 
\begin{equation}
\label{trQQpos}
\mathrm{tr}\left( \mathfrak{Q}(b,\omega)\mathfrak{Q}(b^{\prime},\omega^{\prime})\right)\geq 0 \quad  \forall \, (b,\omega)\ \mathrm{and} \ (b^{\prime},\omega^{\prime})\ \mathrm{a.e.}\,  
\end{equation}
the meaning of the convolution appearing in the integral \eqref{lowsymfsfs}
\begin{equation}
\label{JPGtruedist}
\frac{1}{2\pi} \mathfrak{f}_s\left[\Pi\right]\ast\mathfrak{f}_s\left[\widetilde\Pi\right]
\end{equation}
is clear: it is the probability distribution (w.r.t. the measure $\ud b\,\ud\omega/2\pi$)  for the difference of two vectors in the time-frequency plane, viewed as independent random variables,  and thus is adapted to the abelian and homogeneous structure of the latter (choice of origin  is arbitrary!).   

Let us examine in more details the relationship between $\mathfrak{Q}_0$ and its Weyl transform $\Pi$. First we note the important property issued from \eqref{JPGWTQ}. If $\mathfrak{Q}_0$ is bounded symmetric, i.e. self-adjoint, then its Weyl transform is ``CP'' symmetric:
\begin{equation}
\label{propPipar}
\overline{\Pi(b,\omega)}= \Pi(-b,-\omega)\,. 
\end{equation}
Second, due to the unit trace condition on $\mathfrak{Q}_0$, the symplectic Fourier transform of $\Pi$ is a quasi-probability distribution, 
\begin{equation}
\label{quasiFsPi}
\int_\mathrm{\R^2}  \mathfrak{f}_s\left[\Pi\right](b,\omega) \frac{\mathrm{d} b\, \mathrm{d}\omega}{2\pi}= 1\, .
\end{equation} 
The trivial case  corresponds to   $\Pi(b,\omega) = 1$, of course.  Then $\mathfrak{f}_s\left[\Pi\right](b,\omega)= 2\pi \delta(b,\omega)$ is a probability distribution like its squared convolution  \eqref{trQQpos}, despite the fact that $\mathfrak{Q}_0$ is not a density operator, $\mathfrak{Q}_0 = 2 \mathrm{P}$, where $\mathrm{P}$ is the parity operator (whose the trace  can be consistently defined as $\mathrm{tr}(\mathrm{P})=1/2$).  This no filtering choice yields the popular Weyl-Wigner integral quantization equivalent to the standard ($\sim$ canonical) quantization. 

In the case of Gabor quantization with a probe $\psi$, we already established in \eqref{JPGgabsc1} that \eqref{JPGtruedist} is equal to $ \vert\langle \psi_{b \omega}|\psi_{b^{\prime} \omega^{\prime}}\rangle\vert^2$. The corresponding $\Pi_\psi$ is given by
\begin{equation}
\label{psiPi}
\begin{split}
\Pi_\psi(b,\omega)&= \lg \psi| \mathcal{D}(-b,-\omega)\psi\rg= \lg \mathcal{D}(b,\omega) \psi| \psi\rg\\ &=\sqrt{2\pi}e^{-\ii \frac{b\omega}{2}}\left(\mathcal{F}[\overline{\psi}]\ast\mathcal{F}[\mathrm{t}_{-b}\psi]\right)(\omega)\, , 
\end{split}
\end{equation}
where $ \mathcal{F}$ is the ordinary Fourier transform and $\mathcal{F}[\mathrm{t}_{-b}\psi](\omega)= e^{\ii b\omega}\mathcal{F}[\psi](\omega)$. 
  
We are specially interested in building self-adjoint $A_f$, from physically relevant real functions or distributions $f(b,\omega)$ through \eqref{JPGWHCIQ1}, that also yield  the probability distribution \eqref{JPGtruedist}. An appealing  choice for the operator $\mathfrak{Q}_0$ is that it belongs to the class of density operators for which  \eqref{trQQpos} holds true, or, equivalently,  that the quantity $\mathfrak{f}_s\left[\Pi\right]\ast\mathfrak{f}_s\left[\widetilde{\Pi}\right]/2\pi$, built from  the corresponding $\Pi$, is a probability distribution on the time-frequency plane equipped with the measure $\ud b\,\ud\omega/2\pi$. 
The latter form of the condition is important because  there is no guarantee that  $\mathfrak{f}_s\left[\Pi\right]$ itself is positive and so defines a probability distribution. For instance, in the Gabor case, $\mathfrak{f}_s\left[\Pi_\psi\right]$ is equal to 
 \begin{equation}
\label{FsPiPsi}
\begin{split}
\mathfrak{f}_s[\Pi](b,\omega)&= \int_{-\infty}^{\infty}\ud t \, e^{-\ii \omega t}\,\overline{\psi\left(\frac{t}{2}-b\right)}\, \psi\left(-\frac{t}{2}-b\right)\\ &\equiv 2\pi\mathcal{W}_{\psi}(-b,-\omega)\,,
\end{split}
\end{equation} 
where $\mathcal{W}_{\psi}$ is the Wigner function \cite{wigner32,degosson17}  of $\psi$, and it is well known that  the latter is a probability distribution  only if $\psi(t)$ has the form
\begin{equation}
\label{psigauss}
\psi(t)= \mathrm{Cst}\,e^{-at^2 + bt +c}\,, \quad a,b,c \in \C\,, \quad \mathrm{Re}(a) > 0. 
\end{equation}
Let us now examine the  standard Gaussian examples of  functions $\Pi= \widetilde{\Pi}_G$ having  probabilistic symplectic Fourier transform
  \begin{equation}
\label{Pigauss}
\begin{split}
\Pi_G(b,\omega)= e^{-\frac{b^2}{2\sigma^2}}e^{-\frac{\omega^2}{2\tau^2}}\, , &\quad \mathfrak{f}_s\left[\Pi_G\right](b,\omega)= \sigma\tau e^{-\frac{\tau^2 b^2}{2}}e^{-\frac{\sigma^2\omega^2}{2}}\\
\mathfrak{f}_s\left[\Pi_G\right]^{\ast2}(b,\omega)&= \frac{\sigma\tau}{2} e^{-\frac{\tau^2 b^2}{4}}e^{-\frac{\sigma^2\omega^2}{4}}\, .
\end{split}
\end{equation}

Depending on the pair $(\sigma, \tau)$ the above  functions  do or do not yield non-negative operators  $\mathfrak{Q}_0$. They do  if $\sigma= \tau:= \sqrt{2\tanh \frac{1}{\Theta}}< \sqrt{2}$.  We then obtain the Boltzman-Planck like density operator \cite{JPGber14}:
\begin{equation}
\label{boltplanck}
\mathfrak{Q}_0= \left(1-e^{-\frac{1}{\Theta}}\right)\sum_{n=0}^{\infty}e^{-\frac{n}{\Theta}}|n\rg\lg n| \equiv \rho(\Theta)\, , 
\end{equation}
in the Hilbertian basis made of eigenvectors $|n\rg$ of the harmonic operator 
\begin{equation}
\label{harmop}
\left(\frac{\Omega^2}{2\omega_0^2} + \frac{T^2}{2t_0^2}\right)|n\rg = \frac{1}{\omega_0t_0}\left(n+\frac{1}{2}\right) |n\rg\,.
\end{equation}
For convenience one choose units $\omega_0=1=t_0$. 

In this context, the parameter $\Theta$ plays the role of a temperature (in suitable units). More generally, from the inverse Weyl transform \eqref{invweylQ_0} we obtain  the matrix elements of $\mathfrak{Q}_0$  in the above basis $\{|n\rg\}$ in terms of those of the displacement operator:
\begin{equation}
\label{matelQD}
\left(\mathfrak{Q}_0\right)_{mn}:= \lg m|\mathfrak{Q}_0|n\rg =  \int_{\mathbb{R}^2}  \mathcal{D}_{m n}(b,\omega)\, \Pi(b,\omega) \,\frac{\mathrm{d} b\,\mathrm{d}\omega }{2\pi }\,.
\end{equation} 
The latter involve associated Laguerre polynomials $L^{(\alpha)}_n(t)$ \cite{magnus66}:
\begin{equation}
\label{matelD}
 \mathcal{D}_{m n}(b,\omega) =  \sqrt{\dfrac{n!}{m!}}\,e^{-\vert z\vert^{2}/2}\,z^{m-n} \, L_n^{(m-n)}(\vert z\vert^{2})\, ,   \quad \mbox{for} \ m\geq n\, , 
\end{equation}
with  $L_n^{(m-n)}(u) = \frac{m!}{n!} (-u)^{n-m}L_m^{(n-m)}(u)$ for $n\geq m$, and in which $z:= \frac{1}{\sqrt{2}} (b+\ii \omega)$. Therefore, if the function $\Pi$ is isotropic, i.e. $\Pi(b,\omega):= w(\vert z\vert^2)$, then $\mathfrak{Q}_0$ is the unit-trace diagonal  operator 
\begin{equation}
\label{Q0diagdop}
\mathfrak{Q}_0= \sum_{n=0}^{\infty}\mathcal{L}_n(w) |n\rg\lg n|\, , 
\end{equation}
where
\begin{equation}
\label{lagtr}
\mathcal{L}_n(w)=\left(\mathfrak{Q}_0\right)_{nn} = \int_0^{+\infty}e^{-u/2}\, L_n(u)\,w(u)\,\ud u 
\end{equation}
is a particular Laguerre transform \cite{debnath60,mccully60} with  $$L_n(u)\equiv L^{(0)}_n(u)= \sum_{m=0}^n (-1)^m\binom{n}{m}\frac{u^m}{m!}\,.$$ Since the Laguerre polynomials form an orthonormal basis of the Hilbert space 
$L^2(\R^+, e^{-u}\ud u)$, the transform \eqref{lagtr} is defined for all $n$ and for all $w(u)$ such that $e^{u/2}w(u)$ belongs to this space.  Furthermore, if $w$ is such that $\mathcal{L}_n(w)\geq 0$ for all $n$, then $\mathfrak{Q}_0$ is a density operator. Let us give some hint in the search of solutions to this interesting problem by viewing $w$ as the Laplace transform of a nonnegative function or distribution $\ell$ with support in $[1/2,\infty)$:
\begin{equation}
\label{wellLT}
w(u)= \int_{\frac{1}{2}}^{+\infty}e^{-ut}\,\ell(t)\,\ud t\,. 
\end{equation}
Then from the integral formula \cite{magnus66}, 
\begin{equation}
\label{integasslag}
\begin{split}
\int_{0}^{\infty} \, e^{-\nu x}\, x^{\lambda}\, L_{n}^{\alpha}(x)\ud x &=\frac{\Gamma(\lambda + 1)\Gamma(\alpha+n+1)}{n!\, \Gamma(\alpha + 1)}\nu^{-\lambda -1 }\\
&\times{}_{2}F_1(-n,\lambda + 1; \alpha + 1; \nu^{-1})\, , 
\end{split}
\end{equation}
and assuming that the Frobenius theorem is valid here, we derive
\begin{equation}
\label{ LnTheta}
\mathcal{L}_n(w)=\frac{1}{2} \int_0^{+\infty}e^{-(n+1/2)\Theta}\, \ell(t(\Theta))\,\frac{\ud \Theta}{\Theta^2}\,, \quad t(\Theta)= \frac{1}{2}\coth\left(\frac{1}{2\Theta}\right)\,. 
\end{equation}
There results the convex integral of the Boltzmann-Planck-like density operators $\rho(\Theta)$ defined in \eqref{boltplanck}:
\begin{equation}
\label{Q0ell}
\mathfrak{Q}_0= \frac{1}{2} \int_0^{+\infty} \rho(\Theta)\, e^{-\frac{1}{2\Theta}}\ell(t(\Theta))\,\frac{\ud\Theta}{\Theta^2}\,. 
\end{equation}

\section{Eight-dimensional case and general relativity}
\label{extentGR8}
Hereafter we restrict our quantization procedure to the ones based on the choice of a probe $\psi$. We remind that the Gabor quantization of functions or distributions $f(x,k)$ on the phase space $\{(x,k)\in \mathbb{R}^8\}$ is based on the overcompleteness  \eqref{JPGgabres1} of the family of translated and modulated unit-norm probes  $\psi_{x,k}(y)=e^{\ii k\cdot y}\psi(y-x)$:
\begin{equation}
\label{gabq8}
f(x,k)\mapsto A_f = \frac{1}{(2\pi)^4}
\int_{\R^8} \ud^4 x\,\ud^4 k \,f(x,k)\, \left|
\psi_{x,k}\right\rangle \left\langle \psi_{x,k} \right| \,. 
\end{equation}
 $A_f$ acts  in the Hilbert space of finite-energy signals $s(x) \in \mathcal{H}= L^2(\R^4,\ud^4x)$ as the integral operator
\begin{equation}
\label{ }
s(x)\mapsto (A_f s)(x) = \langle \delta_x  | A_f | s \rangle = 
\int_{\R^4} \ud^4 x^{\prime}\, \mathcal{A}_f(x,x^{\prime})\, s(x^{\prime})\, , 
\end{equation}
with integral kernel given by
\begin{equation}
\label{ }
 \mathcal{A}_f(x,x^{\prime})= \frac{1}{4\pi^2}\int_{\R^4}\mathrm{d}^4 y\, \widehat{f}_{k}(y,x^{\prime}-x)\, \psi(x-y)\,\overline{\psi(x^{\prime}-y)}\,. 
\end{equation}
Here $\widehat{f}_{k}(x,y)$ is the partial Fourier transform with respect to the four-vector variable $k$:
\begin{equation}
\label{partfour4}
\widehat{f}_{k}(x,y)= \frac{1}{4\pi^2}\int_{\R^4}\mathrm{d}^4 k\, f(x,k) \, e^{-\mathsf{i} k\cdot y}\,.
\end{equation}
Remember the non-conventional sign in the Fourier exponential due to our choice of space-time metric. 
The Gabor semi-classical  phase-space portrait of $A_f$ is given by
\begin{equation}
\label{JPGgabsc4} \check{f}(x,k)=  \int_{\mathbb{R}^8}\frac{\mathrm{d}^4 x^{\prime}\,\mathrm{d}^4 k^{\prime}}{(2\pi)^4}\, f(x^{\prime},k^{\prime})\,  \vert\langle \psi_{x,k}|\psi_{x^{\prime}, k^{\prime}}\rangle\vert^2\,.
\end{equation}
Explicit forms of $\check{f}$ are analogous to the time-frequency case \eqref{lowsf} apart from the fact that their writing might become quite cumbersome. 
Here we are essentially concerned  with the quantization and semi-classical portraits of fields on space-time, like the metric field of GR. Therefore the Gabor quantization of the field $u(x)$ yields the multiplication operator  
\begin{equation}
\label{gabqu4}
  (A_{u} s)(x)= \left(u\ast\vert \psi\vert^2\right)(x)\,s(x)\,,
\end{equation}
and its semi-classical portrait is given by
\begin{equation}
\label{lowu4}
\check{u}(x)= \left(u\ast\left(\vert\psi\vert^2\ast\vert\widetilde{\psi}\vert^2\right)\right)(x)=  \left(u\ast R_{\vert\psi\vert^2\vert\psi\vert^2}\right)(x)\,.
\end{equation}
Here convolution and autocorrelations are for functions on $\R^4$ equipped with the its Lebesgue measure $\ud^4 x$: 
\begin{align}
\label{conv4}
 (f\ast g)(x)    & = \int_{\R^4} \ud^4 y \,f(x-y)\,g(y)\,,  \\
\label{autocorrel4}
R_{\psi\psi}(x)&= \int_{\R^4}\mathrm{d}^4 x^{\prime}\,\psi(x^{\prime})\,\overline{\psi(x^{\prime}-x)}=\left(\psi \ast \overline{\tilde{\psi}}\right)(x)\,, 
\end{align}
where we remind that $\tilde{\psi}(x)=\psi(-x)$. 
Repeating \eqref{JPGgabqt} and \eqref{lowtt2}, the space-time variables acquire the status of quantum observables acting on the space of signals:
\begin{align}
\label{JPGgabx4}
 x^{\mu}  \mapsto  A_{x^{\mu}} &= X^{\mu}   -\lg X^{\mu} \rg_{\psi} \,\mathbbm{1} \,, \quad X^{\mu} (s)(x)= x^{\mu} s(x)\,,\\
 \label{lowx4}
\check{x^{\mu} }&= x^{\mu}  - \lg x^{\mu} \rg_{R_{\vert\psi\vert^2\vert\psi\vert^2}}\,. 
\end{align}
With the choice of the Gaussian probe
\begin{equation}
\label{gaussp4}
G_{\pmb\sigma}(x)= \prod_{\mu=0,1,2,3} G_{\sigma_{\mu}}(x^{\mu})\, ,
\end{equation}
the above formula assume the simple expressions:
\begin{align}
\label{gabqgauss4}
   (A_{u} s)(x) &= \left(u\ast G^2_{\pmb\sigma}\right)(x)\,s(x)\,, \quad  \sqrt{\pmb\sigma}:= \prod_{\mu=0,1,2,3} \sqrt{\sigma_{\mu}} \,,   \\
  \label{gabscgauss4}  \check{u}(x)&=  \left(u\ast \left(G^2_{\pmb\sigma}\ast G^2_{\pmb\sigma}\right) \right)(x) =   \left(u\ast G^2_{\sqrt{2}\pmb\sigma} \right)(x) \,.  
  \end{align} 
The above formalism allows us to give a quantum version of the metric fields $(g_{\mu\nu}(x))$ of  general relativity,
\begin{equation}
\label{qgmn}
g_{\mu\nu}(x) \mapsto A_{g_{\mu\nu}}\,, 
\end{equation}
and to give a regularised version 
depending of the choice of probe,
\begin{equation}
\label{lwgmn}
\check{g}_{\mu\nu}(x)\,. 
\end{equation}
As a first example, let us examine the Minkowskian metric in Cartesian coordinates, $(g_{\mu\nu})= \mathrm{diag}(1,-1,-1,-1)$. We trivially get $\left(A_{g_{\mu\nu}}\right)= \mathrm{diag}(\bu,-\bu,-\bu,-\bu)$. Let us now examine what happens if one quantizes the same  metric as is expressed in cylindric coordinates $(g^{\mathrm{cyl}}_{\mu\nu}(x^0,\rho,\theta,x^3)= \mathrm{diag}(1,-1,-\rho^2,-1)$. This metric is singular in the sense that its determinant  cancels on the third axis $\rho=0$. Due to the rotational symmetry about this axis, we naturally choose $\pmb\sigma=(\sigma_0, \sigma, \sigma, \sigma_3)$. We then find the new metric operator and its semi-classical portrait:
\begin{equation}
\label{cylmet}
\begin{split}
\left(A_{g_{\mu\nu}}\right)& = \mathrm{diag}(\bu,-\bu,-(\hat{\rho}^2 + \sigma^2\bu),-\bu)\, , \quad \hat{\rho}^2= \left(X^1\right)^2 +  \left(X^2\right)^2\, , \\
\ud s^2 &= \ud {x_0}^2 - \ud\rho^2  - (\rho^2 + 2\sigma^2)\ud\theta^2 - \ud {x_3}^2\,. 
\end{split}
\end{equation}
Both expressions have become regular on the third axis.
  
 In the case of isotropic space  spherical coordinates for the spatial part are privileged, i.e., $(g^{\mathrm{spher}}_{\mu\nu}(x^0,r,\theta,\phi)= \mathrm{diag}(1,-1,-r^2,-r^2\sin^2\theta)$. This metric is singular as its determinant  cancels at the origin $r=0$. Choosing $\pmb\sigma=(\sigma_0, \sigma, \sigma, \sigma)$. 
we obtain for the metric operator and  its semi-classical portrait:
\begin{equation}
\label{sphmet}
\begin{split}
\left(A_{g_{\mu\nu}}\right)& = \mathrm{diag}\left(\bu,-\bu,-\left(\hat{r}^2 + \frac{3}{2}\sigma^2\bu\right),-(\hat{\rho}^2 + \sigma^2)\bu\right)\, ,  \\
\ud s^2 &= \ud {x_0}^2 - \ud r^2  - \left(r^2 + 3\sigma^2\right)\ud\theta^2 - (\rho^2 + 2\sigma^2)\ud \phi^2\,, 
\end{split}
\end{equation}
with $\hat{\rho}^2= \left(X^1\right)^2 +  \left(X^2\right)^2$.  Both expressions are now regular at the origin.

Be aware that simply adding a positive constant to $\rho^2$ (cylindrical case) and to $\rho^2$ and $r^2$ (spherical case) creates non-trivial changes to the Euclidean geometry of the spatial part of the metric. One shows that deformations of geodesics from straight lines involve elliptic functions  \cite{gaczu22}. 

Next, if the metric fields are known solutions of the Einstein equations for a given tensor energy density \begin{equation}
\label{einssource}
R_{\mu\nu}- \frac{1}{2}R= \kappa\,T_{\mu\nu}\, , \quad \kappa= \frac{8\pi G}{c^4}\,,
\end{equation}
then their respective $\psi$ regularised versions give rise to the modified tensor energy density $\check{T}_{\mu\nu}$ through the equations:
\begin{equation}
\label{einssource}
\check{T}_{\mu\nu}=  \frac{1}{\kappa}\left(\check{R}_{\mu\nu}- \frac{1}{2}\check{R}\right) \,. 
\end{equation}
This procedure offers the opportunity to work with smooth versions of the metric field, where the smoothing has a probabilistic content. 
Moreover, the notion of empty space becomes  mathematically totally idealistic, since any quantization of the above Gabor type introduces (fictitious? real?) matter,  however minute it may be. This matter arises from the lack of information which is encoded in the parameters of the probe $\psi$. 

 The important point to be examined is the physical meaning of the set of parameters $\boldsymbol{\kappa}$ introduced in Section \ref{JPGsigtoqu}, or in more illustrative terms, the set  $\pmb\sigma$ of gaussian widths introduced in   \eqref{gaussp4}.  The probabilistic nature of these parameters should reflect our inability to reach exactness in terms of the information about the observed system through data and interpretative model. There exist limit values assumed by  these parameters beyond which the mathematical model loses its physical relevance, i.e. where no measurement or observation makes sense. Those limit values should be viewed as based on  or originate in the existence of a unique quantum of information, existing besides  other elementary quanta as they were  listed by one of us in \cite{GCT18,GCT19}. It would be fruitful to revisit the present approach from a more informational viewpoint. See for instance \cite{djafari15}  for a survey of the employed concepts, and  \cite{souriau78,marle20A,marle20B,barbagay20} for a more specific \`a la Souriau viewpoint based on the symmetry of considered manifolds, the simplest one being the phase space $\R^8$. We already tackled the subject with the expressions \eqref{boltplanck}  and \eqref{Q0ell}.

\section{The example of the uniformly accelerated reference system}
\label{unac}
In their illuminating 1935 article \cite{einstein-rosen35}, Einstein and Rosen state that 
\begin{quote}
\textit{Every field, in our opinion, must therefore adhere to the fundamental principle that singularities of the field are to be excluded.}
\end{quote}
Needless to say that we agree with this statement, and the first example they provide gives us the opportunity to show how  Gabor quantization of the metric field   cures the singularity problem. Following the Principle of Equivalence one can view  a uniformly  accelerated reference system as at rest provided that there exists a homogeneous gravitational field in space with respect to it.  The corresponding metric field is given by
\begin{equation}
\label{unifa}
\ud s^2= \alpha^2 x_1^2 \ud x_0^2 - \ud x_1^2 - \ud x_2^2- \ud x_3^2\, . 
\end{equation} 
The singularity holds on the hyperplane $x_1=0$.
For $x_1\neq 0$ this leads to the Einstein field equations  
\begin{equation}
\label{unifaeq}
R_{\mu\nu}=0\,.
\end{equation} 
The restriction $x_1\neq 0$ is necessary since the Ricci tensor is indeterminate on the hyperplane $x_1=0$. In view of regularisation the authors of \cite{einstein-rosen35}  propose to modify the metric \eqref{unifa} with the introduction of a small constant $\varsigma$ as
\begin{equation}
\label{unifareg}
\ud s^2= \left(\alpha^2 x_1^2  + \varsigma\right)\ud x_0^2 - \ud x_1^2 - \ud x_2^2- \ud x_3^2\, . 
\end{equation} 
This ansatz gives rise to a regular metric  and a new stress-energy tensor $\check{T}_{\mu\nu}$  through the gravitational  equation with source
\begin{equation}
\label{fictsource}
\check{R}_{\mu\nu}- \frac{1}{2}g_{\mu\nu}\,\check{R}= \kappa\,\check{T}_{\mu\nu}\, , \quad \kappa= \frac{8\pi G}{c^4}\,,
\end{equation}
whose the nonvanishing components read
\begin{equation}
\label{n0comp}
\check{T}_{22}=\check{T}_{33}=- \frac{\alpha^2\varsigma}{\kappa} \left(\alpha^2x_1^2 + \varsigma\right)^{-2} \,, 
\end{equation}
which cancels at $\varsigma=0$, as expected. This fictitious or not stress-energy tensor is a pure pressure in the directions 2 and 3. 

Now the Einstein-Rosen ansatz is not  an ansatz any longer if we apply the Gabor quantization procedure to the metric field in \eqref{unifa} and the resulting semi-classical portraits. This is the reason why we checked  the quantities appearing in \eqref{fictsource}. To simplify we choose the Gaussian probe \eqref{gaussp4}. 
From \eqref{gauss2qt2} we get for the metric operators:
\begin{equation}
\label{qmetunif}
\left(A_{g_{\mu\nu}}\right)= \mathrm{diag}\left(
  \alpha^2 X_1^2 +  \alpha^2\frac{\sigma_1^2}{2} \mathbbm{1}, - \mathbbm{1},- \mathbbm{1},- \mathbbm{1}\right)\, , \quad X_1s(x)=x_1s(x)\, , 
\end{equation}  
and for their semi-classical portrait, 
\begin{equation}
\label{scmetunif}
\left(\check{g}_{\mu\nu}\right)= \mathrm{diag}\left(
  \alpha^2 x_1^2 +  \alpha^2\sigma_1^2, -1,-1,-1\right)\, .
\end{equation}  
Thus the Einstein-Rosen parameter can be identified as 
\begin{equation}
\label{anssigma}
\varsigma= \alpha^2\sigma_1^2
\end{equation}
and is interpreted as proportional to the Gaussian variance  for the variable $x_1$.  Hence, \eqref{n0comp} becomes
\begin{equation}
\label{n0comp}
\check{T}_{22}=\check{T}_{33}=- \frac{\sigma_1^2}{\kappa} \left(x_1^2 + \sigma_1^2\right)^{-2} \,, 
\end{equation}
The regularisation parameter $\sigma_1$  conveys our  degree of ignorance of what is the ``exact'' geometric modelling of the above homogeneous gravitational field in space. 

\section{The example of the Schwarzschild metric field}
\label{szchw}
We now turn our attention to the Schwarzschild solution for the static spherically symmetric field produced by a spherical symmetric body at rest:
\begin{equation}
\label{szchwmet}
\begin{split}
\ud s^2 &= \left(1-\frac{2m}{r}\right)\ud t^2  -\left(1-\frac{2m}{r}\right)^{-1}\ud r^2 -r^2\ud \theta^2 -r^2\sin^2\theta \ud\phi^2
\\ & \equiv U\ud t^2 -V\ud r^2 -r^2\ud \theta^2 -r^2\sin^2\theta \ud\phi^2 \,,
\end{split}
 \end{equation}
with appropriate units. Note that the  metric for the de Sitter space-time with horizon has the same form, just replace $2m/r$ with $\Lambda r^2/3$ in $g_{00}$ and $g_{rr}$, where  $\Lambda$ is the positive cosmological constant.  The cosmological horizon holds at $r= \sqrt{3/\Lambda}$.

Since the Schwarzschild solution holds outside the surface of the body that is producing the field, i.e. for $2m<r<\infty$, and since $r=0$ and $r=2m$ are singularities, pseudo for the latter  since the determinant of the metric does not cancel at $r=2m$,   we have to examine if our approach is able to regularise these singularities to a certain extent.  
Let us pick a regular isotropic $\psi$ such that $\vert \psi \vert^2(x)\equiv \vert \psi \vert^2(t,r)$ and, consequently, $R_{\vert\psi\vert^2\vert\psi\vert^2}(r)$ are  probability distributions. Applying \eqref{gabqu4} and \eqref{lowu4} yields the metric multiplication operators 
\begin{align}
\nonumber 
  (A_{U} s)(x)&= \left(U\ast\vert \psi\vert^2\right)(x)\,s(x)= \left(1-\frac{2m}{r} +2m\left\lg Y_{r}(r^{\prime})\left(\frac{1}{r}-\frac{1}{r^{\prime}}\right)\right\rg_{\vert \psi \vert^2}\right)s(x)\\
 \label{gabqU} &\equiv  \widetilde{U}_{\vert \psi\vert^2} (r)\,s(x)\\
 \nonumber (A_{V} s)(x)&= \left(V\ast\vert \psi\vert^2\right)(x)\,s(x)= \left(1+\frac{2m}{r} + \frac{2m}{r}\left\lg\frac{m}{r^{\prime}}\ln\frac{\vert r+r^{\prime}-2m\vert}{\vert r-r^{\prime}-2m\vert}\right\rg_{\vert \psi \vert^2} +\right.\\
 \nonumber  &  \left. + 2m\left\lg Y_{r}(r^{\prime})\left(\frac{1}{r^{\prime}}-\frac{1}{r}+\frac{m}{rr^{\prime}}\ln\frac{\vert r-r^{\prime}-2m\vert}{\vert r^{\prime}-r-2m\vert}\right)\right\rg_{\vert \psi \vert^2}\right)s(x)\\
 \label{gabqV}  &\equiv  \widetilde{V}_{\vert \psi\vert^2}(r)\,s(x)\equiv \left(2- \widetilde{U}_{\vert \psi\vert^2} (r) + L_{\vert \psi\vert^2}(r)\right)\,s(x)\\
 \label{gabqr2}   (A_{r^2} s)(x)&= \left(r^2\ast\vert \psi\vert^2\right)(x)\,s(x)= \left(r^2 + \lg {r^{\prime}}^2\ \rg_{\vert \psi \vert^2}\right)s(x)\,,\\
   \label{gabqrs2}  (A_{r^2\sin^2\theta} s)(x)&= \left(r^2\sin^2\theta\ast\vert \psi\vert^2\right)(x)\,s(x)= \left(r^2\sin^2\theta + \frac{2}{3}\lg {r^{\prime}}^2\ \rg_{\vert \psi \vert^2}\right) s(x)\,,
\end{align}
after regularisation of  logarithmic singularity  through Cauchy principal value in \eqref{gabqV}.  We have primed the integration variable in the computation of expected values in order to avoid the confusion with the external variable $r=\Vert \pmb x \Vert$, and we have introduced   the Heaviside function: $Y_{r}(r^{\prime}):= Y(r^{\prime}-r)$. Note that the $2/3$ in \eqref{gabqrs2} is the average of $\sin^2\theta$ on the sphere.   Remind that the notation  $\lg f \rg_{p}$ stands for the expected value of the function $f$ with respect to the probability distribution $p$. 

The corresponding  semi-classical portraits have similar expressions provided that we replace $\vert \psi \vert^2$ with $R_{\vert\psi\vert^2\vert\psi\vert^2}$:
\begin{align}
\label{scgabqU}
  \check{U}(x)&= \left(U\ast R_{\vert\psi\vert^2\vert\psi\vert^2}\right)(x)=\widetilde{U}_{R_{\vert\psi\vert^2\vert\psi\vert^2}}(r)\,,\\
 \label{scgabqV}  \check{V}(x)&= \left(V\ast R_{\vert\psi\vert^2\vert\psi\vert^2}\right)(x)= \widetilde{V}_{R_{\vert\psi\vert^2\vert\psi\vert^2}}(r)
 \,,\\
 \label{scgabqr2}    \check{r^2}(x)&= \left(r^2\ast R_{\vert\psi\vert^2\vert\psi\vert^2}\right)(x)= r^2 + \lg {r^{\prime}}^2\ \rg_{R_{\vert\psi\vert^2\vert\psi\vert^2}}\,,\\
   \label{scgabqrs2}   \check{r^2\sin^2\theta}(x)&= \left(r^2\sin^2\theta\ast\R_{\vert\psi\vert^2\vert\psi\vert^2}\right)(x)= r^2\sin^2\theta + \frac{2}{3}\lg {r^{\prime}}^2\ \rg_{R_{\vert\psi\vert^2\vert\psi\vert^2}}\,,
\end{align}

Let us  analyse  the functions $\widetilde{U}_{p}(r)$ and $\widetilde{V}_{p}(r)$, where $p=p(t,r)$  denotes  a spatially isotropic probability distribution  on $\R^4$, as are $\vert\psi\vert^2$ or $R_{\vert\psi\vert^2\vert\psi\vert^2}$. 
Let us write their expressions after rearranging some terms:
\begin{align}
\label{UTpr}
  \widetilde{U}_{p}(r)  &=  1-\frac{2m}{r} +2m\left\lg Y_{r}(r^{\prime})\left(\frac{1}{r}-\frac{1}{r^{\prime}}\right)\right\rg_{p}= 1 - \frac{2m}{r} \left\lg \mathds{1}_{[0,r]}(r^{\prime})\right\rg_{p}-2m\left\lg Y_{r}(r^{\prime})\frac{1}{r^{\prime}}\right\rg_{p}\, ,   \\
  \label{VTpr} \widetilde{V}_{p}(r)  &=  2- \widetilde{U}_{p} (r)  + L_p(r)\, , 
\end{align}
with 
\begin{align}
\label{Lpr1}
L_p(r)&= \frac{2m^2}{r}\left\lg\frac{1}{r^{\prime}}\left[\ln\frac{\vert r^{\prime} +(r-2m)\vert}{\vert r^{\prime}-(r-2m)\vert} + Y_{r}(r^{\prime})\ln\frac{\vert r^{\prime}-(r-2m)\vert}{\vert r^{\prime}-(r+2m)\vert}\right]\right\rg_{p}\\
\label{Lpr2}
&=\frac{2m^2}{r}\left\lg\frac{1}{r^{\prime}}\left[\mathds{1}_{[0,r]}(r^{\prime})\ln\frac{\vert r^{\prime}+ (r-2m)\vert}{\vert r^{\prime}- (r-2m)\vert}  +  Y_{r}(r^{\prime})\ln\frac{\vert r^{\prime}+(r-2m)\vert}{\vert r^{\prime}-(r+2m)\vert}\right]\right\rg_{p}
\end{align}

\subsection*{Regularisation at  Schwarzschild radius value}
We note the regularisation of the metric at the classical singularity $r=2m$:
\begin{align}
\label{regulszchwU}
\widetilde{U}_{p}(2m)  &=  \left\lg Y_{2m}(r^{\prime})\left(1-\frac{2m}{r^{\prime}}\right)\right\rg_{p}>0\, , \\
 \label{regulszchwV} \widetilde{V}_{p}(2m) &=  1  + \left\lg \mathds{1}_{[0,2m]}(r^{\prime})\right\rg_p + \left\lg Y_{2m}(r^{\prime}) \left(\frac{2m}{r^{\prime}} +\frac{m}{2r^{\prime}}\ln\frac{ r^{\prime}}{ \vert r^{\prime}-4m\vert}\right)\right\rg_{p} > 0\,.
\end{align}

\subsection*{Behaviour at large $r$}
Trivially  we have 
\begin{equation}
\label{behinf}
\widetilde{U}_{p}(r) \underset{r\to \infty}{\to} 1\, , \quad \widetilde{V}_{p}(r) \underset{r\to \infty}{\to} 1\,.
\end{equation}
\subsection*{Behaviour at  $r=0$}
Less trivially, we have
\begin{equation}
\label{beh0}
\widetilde{U}_{p}(r) \underset{r\to 0}{\to} 1 -\left\lg \frac{2m}{r^{\prime}}\right\rg_{p}\, , \quad \widetilde{V}_{p}(r) \underset{r\to 0}{\to} 1 +\left\lg \frac{2m}{r^{\prime}}\right\rg_{p} + 4m^2 \left\lg \frac{1}{r^{\prime}(r^{\prime}-2m)}\right\rg_{p}\,, 
\end{equation}
where the last mean value should be understood in the sense of Cauchy principal value, more precisely  Hilbert transform,  for the integral 
\begin{equation}
\label{Hilbtrrp}
\left\lg \frac{1}{r^{\prime}(r^{\prime}-2m}\right\rg_{p}= 4\pi\int_{-\infty}^{+\infty}\ud t^{\prime}\int_0^{+\infty}\ud r^{\prime} \frac{r^{\prime} p(t^{\prime}, r^{\prime})}{r^{\prime}-2m}= -4\pi^2\int_{-\infty}^{+\infty}\ud t^{\prime}\,\operatorname {H}(Y(\cdot)(\cdot)p(t^{\prime}, (\cdot))(2m)\,, 
\end{equation}
with the notation
\begin{equation}
\label{Hilbtr}
\displaystyle \operatorname {H} (u(\cdot))(t)={\frac {1}{\pi }}\,\operatorname {p.v.} \int _{-\infty }^{+\infty }{\frac {u(\tau )}{t-\tau }}\;\mathrm {d} \tau\,.
\end{equation}
Now, the positiveness of the derivative of $\widetilde{U}_{p}$, 
\begin{equation}
\label{dUpr}
\frac{\ud \widetilde{U}_{p}}{\ud r}(r)= \frac{2m}{r^2}\,\left\lg \mathds{1}_{[0,r]}(r^{\prime})\right\rg_{p}> 0\, ,
\end{equation}
 entails that $\widetilde{U}_{p}(r)$ is monotone increasing from $U_{\mathrm{min}}=1 -\left\lg \dfrac{2m}{r^{\prime}}\right\rg_{p}$ (at which the slope is infinite) to $1$. Hence one deals with two cases. 
\begin{enumerate}
  \item $U_{\mathrm{min}}> 0$, i.e., $\left\lg \dfrac{2m}{r^{\prime}}\right\rg_{p}< 1$. Then the temporal term of the  Schwarzschild metric is completely regularised. 
  \item $U_{\mathrm{min}}\leq 0$, i.e., $\left\lg \dfrac{2m}{r^{\prime}}\right\rg_{p}\geq 1$. Then there exists a smaller ``Schwarzschild radius" $r_{s0}\in (0,2m)$ for the temporal part, defined by the equation
  \begin{equation}
\label{TszchR}
r_{s0}=  \frac{2m \left\lg \mathds{1}_{[0,r_{s0}]}(r^{\prime})\right\rg_{p}}{1-2m\left\lg Y_{r_{s0}}(r^{\prime})\frac{1}{r^{\prime}}\right\rg_{p}}\, . 
\end{equation} 
\end{enumerate} 
The situation is less obvious for the radial metric term $\widetilde{V}_{p}(r)$. 
Nevertheless, through a suitable choice of probability distribution one can expect a full regularisation too.

A more developed analysis with explicit examples together with the determination of the geodesics and of the stress-energy tensor $\check{T}_{\mu\nu}$ resulting of this regularisation and its interpretation will be addressed in a forthcoming work. 

\section{Final discussion}
\label{findis} 

Let us initiate the discussion by quoting  the authors of \cite{kuncas20}: 
\begin{quote}
\textit{... infinities in General relativity come in the form of singularities \cite{hawkell73} and they point to the breakdown of our current understanding of gravity. The standard view in the community is that quantum gravity should be able to resolve this issue by smoothing out singularities. Nonetheless, despite the many existing approaches to quantum gravity, there is no consensus about what it is and how one should construct a quantum theory of spacetime, thus a proof of principle for the singularity avoidance is yet to be found.
}
\end{quote}

The present work should be viewed as a partial contribution to this program.  We have presented a  non-orthodox quantization which transforms functions on the eight-dimensional phase space (time-space, frequency-wavevector) into operators in the Hilbert space of signals on space-time, or equivalently, in the Hilbert space of their Fourier transforms. Non-orthodox means that there is no Planck constant involved in our approach, and so there is no associated particle, no mass, no energy, no momentum associated with the resulting quantum objects. We remind that with the Planck constant  one enters the domain of quantum mechanics and quantum field theory with its Fock space formalism. Its fundamental role is to  bridge two worlds, for instance the world of classical waves  with its phase space (time-space, frequency-wave vector)   and the quantum mechanics built from the classical phase space (space, momentum).

As a first exploration, we have considered the metric field of general relativity, but we could as well examine the Maxwell  electromagnetic field. These fields are functions of space-time coordinates, but nothing prevents us to extend their variable domain with the inclusion of frequencies and wave vectors. The latter could be viewed as phenomenological quantities, analogous to those introduced in various studies, e.g., waves in plasmas.

Concerning  general relativity, we have shown that the quantization of a metric field that is solution of the Einstein equation in empty space gives rise to some regularised version of it  and to a stress-energy tensor. Although the latter could be  thought as ``fictitious'', we believe on  the contrary that its existence is unescapable due to our ignorance of an ``exact'' mathematical model for describing gravitation. In this regard,  non-emptiness of space-time is a type of attribute which might appear to physicists as idealistic or superfluous while it is unescapable  as a completion of any mathematical model. 

Finally, one could think that our approach of transforming space-time points and  metric fields into operators  is parent of non-commutative geometry, as is nicely presented,  for instance, in \cite{connes07}. Actually, our quantization of the classical geometry of space-time yields a (formal) commutative algebra of multiplication operators, formal meaning that we make abstraction of domain considerations in dealing with products of such operators. On the other hand,  the algebra becomes non-commutative  if we quantize extensions of the metric field to functions depending on frequency and wave-vector.

\end{document}